# Confronting Machine Learning With Financial Research

Kristof Lommers, Ouns El Harzli, Jack Kim[1]

**Abstract**

This study aims to examine the challenges and applications of machine learning for financial research. Machine learning algorithms have been developed for certain data environments which substantially differ from the one we encounter in finance. Not only do difficulties arise due to some of the idiosyncrasies of financial markets, there is a fundamental tension between the underlying paradigm of machine learning and the research philosophy in financial economics. Given the peculiar features of financial markets and the empirical framework within social science, various adjustments have to be made to the conventional machine learning methodology. We discuss some of the main challenges of machine learning in finance and examine how these could be accounted for. Despite some of the challenges, we argue that machine learning could be unified with financial research to become a robust complement to the econometrician's toolbox. Moreover, we discuss the various applications of machine learning in the research process such as estimation, empirical discovery, testing, causal inference and prediction.

---

[1] Kristof Lommers (University of Oxford Said Business School and Oxford-Man Institute of Quantitative Finance), Jack Kim (Data Capital Management), Ouns El Harzli (University of Oxford). Correspondence via Kristof Lommers (mail: Kristof.Lommers.DPHIL@said.oxford.edu). We would like to thank Viktor Smits, Stan Beckers, and Lina Thomas for their helpful comments on the paper.

# 1. Introduction

During the past decade, machine learning has gained substantial interest among scientists but its adoption in financial-economic academia has been rather limited. Given its success in other fields and the empirical nature of finance, it is appealing to use machine learning for the purpose of financial research. It is important to keep in mind that the models themselves are biased as they optimize a restricted objective function according to a specific algorithmic methodology and statistical rationale. As a result, one should be cautionary about the application of machine learning in financial markets which present an unique data environment and scientific context. Finance is, despite its quantitative character, a social science where the underlying research paradigm differs considerably from other fields in which most machine learning progress has been made. We attempt to provide a broad discussion on the application of machine learning in financial-economic research and put emphasis on the peculiarities of the financial market environment. We discuss the particular challenges of machine learning for financial research and examine how to appropriately adjust the methodological framework to address these. We argue that machine learning poses promising opportunities when applying appropriate adjustments to fit the idiosyncrasies of financial markets. Furthermore, we discuss how machine learning could be reconciled with the scientific research paradigm in finance. More specifically, we argue that despite the large differences between econometrics and machine learning, they could be reconciled into a broader framework of financial machine learning. Contrary to the popular view that machine learning is a black box that is only useful for prediction, we argue that it can be an intuitive tool to help researchers to construct theory in the face of the complex realities and data structure of financial markets. Machine learning has various applications within the empirical research process such as estimation, empirical discovery, testing, causality inference, and prediction. Various papers have argued in favour of the promising opportunities that machine learning provides to research in finance and economics. Varian (2014) asserts that economists should use machine learning in order to computationally handle the size, unstructured nature and high dimensionality of modern data sources. Mullainathan and Spiess (2017) contend that machine learning provides improved tools for the purpose of prediction and estimation in economics. Gu, Kelly and Xiu (2020) argue that machine learning methods can be especially helpful in empirical asset pricing as risk premium measurement fundamentally constitutes a prediction task. Machine learning would allow for more functional flexibility in estimating non-linear risk premia and accommodate more potential predictors as it can handle higher dimensional data. Stevanovic and Surprenant (2019) provide a similar argument, and observe that machine learning improves macroeconomic predictions by capturing non-linearities that are present as a result of frictions and inefficiencies. Athey and Imbens (2017) convey that machine learning allows economists to enhance their empirical toolbox with more sophisticated techniques and improve cross-disciplinary communication with other researchers. This paper contributes to the literature on two fronts. First, we discuss how machine learning models fit the idiosyncrasies of financial data and examine how machine learning methodologies could be adjusted for this. The sub-field of financial machine learning is in its infancy within the academic literature, and discussing how machine learning fits the challenges and specificities of financial data is an important step towards establishing this field. Second, we examine how machine learning fits the paradigm of empirical research in finance. Whether machine learning will be taken seriously depends on how well it fits with how economists think and whether it could be reconciled with econometrics. We assert that machine learning poses several opportunities to enhance the empirical framework used by financial economists and should be viewed as a complement to the econometrician's toolbox.

## 2. Confronting Machine Learning with the Financial Research Paradigm

In defining machine learning we follow Molina and Garip (2019) who consider it as an "offshoot" of nonparametric statistics which lies at the intersection of statistical learning, optimization and computer science. Machine learning algorithms essentially encapsulate a statistical optimization problem with a loss function, an optimization criterion, and an optimization routine as fundamental building blocks (Burkov, 2019). The large variety of learning algorithms essentially differ in the specification of these three building blocks and the degree of supervision given to the algorithm. Research in finance generally involves the use of applied statistics, mostly in the form of regression analysis, to examine relationships. In this regard one would expect that machine learning would be an excellent fit for finance. Coulombe, Leroux, Athey and Imbens (2019) discuss why machine learning has not penetrated economic academia yet as a result of a clashing culture. There are three main conflicts between financial-economic research and machine learning relating to the importance of statistical inference, causality, and a priori hypotheses and model assumptions.

First, empirical research in finance and economics has a strong focus on statistical inference. Econometrics has been built upon the rationale of statistical inference and it has become the way economists think about empirical research. The main difference between econometrics and machine learning could be found in the focus on inference relative to prediction (Bzdok, Altman and Krzywinski, 2018). Prediction focuses on the prediction output, its quantitative accuracy and generalizability while statistical inference focuses on learning information about the structure of the underlying data generating process and the relationships between variables (Sanders, 2019; Bzdok, Altman, and Krzywinski, 2018). In inference one wants to characterize the underlying data generating process in the form of a probability model and infer a range of statistical measures based on this model. The a priori specification of a probability model, that is proven to be theoretically sound under certain assumptions, allows to measure uncertainty and statistically infer some characteristics of the underlying data generating process. Economists generally put a lot of attention on parameter estimation and the tools within inference have been a theoretically reliable methodology to achieve this (Mullainathan and Spies, 2017). There is a strong link with probability theory as one makes assumptions about the distribution where these are strongly linked to a limited set of "canonical" models within econometrics (Biddle, 2016). One could argue that most traditional econometric model specifications are primarily motivated by amenability to statistical inference. Most traditional econometric models are endowed with some form of conditional linearity, additivity, and monotonicity. These generative formulations describing the relationship between innovations of covariates lead to a purported ease of testing the statistical significance of underlying hypotheses via parameter p-values. In machine learning there is a focus on prediction where one tries to algorithmically optimize models to fit the underlying data generating process as well as possible. In machine learning, models are not assumed to define the underlying data generating process but are considered as a means to an end for the purpose of prediction. If the model performs well in-sample, it is likely to generalize well and no further assumptions on the data generating process are necessary. As discussed by de Prado (2019) and Mullainathan and Spies (2017), the rationale behind the traditional econometric approach is not compatible with out-of-sample prediction. A large part of the machine learning literature focuses on predictive learning and there is a general emphasis on predictive performance. Machine learning focuses on out-of-sample prediction quality because when the model describes the data well it should generalize equally well and little assumptions on the data generating process are needed (Rudin, 2015). It should be noted that in some areas of finance, such as asset pricing, prediction is an important component. Asset pricing essentially encompasses the estimation of risk premia, which represent expected returns of risky assets, and thus involves a substantial prediction component (Gu, Kelly and Xiu, 2020). The process of economically interpretable feature importance will lie at the heart of machine learning as an empirical inference tool as opposed to solely a tool for prediction. How machine learning could be adjusted to fit the rationale behind inference and various advancements within feature importance have been made during the past years.

Second, causal understanding has a central place in financial economics while most machine learning methods do not place much emphasis on causation (Rudin, 2015). In financial research one wants to fundamentally understand the relationships and the dynamics causing them. For this reason economists have used econometrics and empirical design methods such as instrumental variables, difference-in-differences and regression discontinuity (Varian, 2014). Machine learning does not generally focus on understanding of the relationships but tries to learn the underlying structure and predict the relationships as best as possible. As a result, the most widely used machine learning algorithms have limited explanatory power as the focus is on prediction performance (Jung, Patnam and Ter-Martirosyan, 2018). At first sight it seems difficult to bring machine learning and financial research together when the core goal seems to be fundamentally different. However, machine learning techniques can be useful for causal inference and substantial progress made in this field. As discussed later in this paper, machine learning could be used for causal inference in two ways, namely, to improve traditional methods in causal inference via improved prediction and to introduce new methodologies for causality analysis. Since the analysis of causality is an arduous task with traditional methods, the growing sub-field of machine learning tackling causality could be an useful addition to the financial researcher's toolbox.

Third, financial economics is hypothesis-driven while machine learning tends to be data-driven (Rudin, 2015). More specifically, financial models rely on pre-specified hypothesized relationships as researchers assume a certain functional form governing them. This is linked to the aforementioned argument on statistical inference and simplified functional forms are used to allow for theoretically sound results allowing for inference. Machine learning models, on the other hand, do not make many a priori assumptions on the underlying relationships as the focus is on algorithmically finding a functional form that describes the relation between the variables (Jung, Patnam and Ter-Martirosyan, 2018). It should be noted that machine learning algorithms do make certain data and relationship assumptions, but these tend to be significantly less stringent relative to the ones used in conventional econometric methods. This principle is supported by the no-free-lunch theorem which states that there are no a priori differences in performance between various machine learning algorithms when averaged across problems (Wolper and Macready, 2005). One of the core paradigms behind machine learning is thus the flexibility to fit the underlying structure of the data as best as possible while trying to generalize well and control for overfit. However, it should be noted that each machine learning algorithm is a model with its own bias and appropriate use cases. Machine learning focuses more on learning from the data and the specific functional form of the relationship is decided based on the best fit with the underlying data. In financial research one a priori specifies the functional specification governing the variables within the econometric model while in machine learning algorithms learn the specifications from the data (Mullainathan and Spiess, 2017). The hypothesis-driven approach has long dominated scientific research, however, has been declining in the era of broadly available computing power and data. We would like to argue that researchers should aspire for a balance between the two as both approaches do not necessarily supplement one another but could complement each other. Later in this paper, we discuss approaches where machine learning could be used for a methodology that combines the data-driven and hypothesis-driven rationale for scientific research.

Given the aforementioned elements one could infer that machine learning traditionally solves a different problem than the one econometrics attempts to tackle (Mullainathan and Spies, 2017). As discussed further in this paper, machine learning could be applied to improve various parts of the empirical research process such as data preprocessing, empirical discovery, estimation, testing, prediction and causality analysis. Machine learning is not a supplement to econometrics but rather a complement to the econometric toolbox. As argued by Bzdok, Altman and Krzywinski (2018) machine learning and methods of statistical inference are complementary in finding meaningful relationships in the data. Moreover, as argued by Athey and Imbens (2017), it could allow financial economists to not only enhance their empirical toolbox with more sophisticated techniques but also improve cross-disciplinary communication with other researchers who increasingly embrace machine learning as a scientific tool. Despite many promising use cases, there is still substantial progress to be made in the cross-section of econometrics and machine learning.

## 3. Addressing the Challenges of Machine Learning in Finance

It is important to keep in mind that each machine learning algorithm is its own model with its specific assumptions and underlying algorithmic rationale. Given that machine learning essentially is a statistical optimization framework, its algorithms are based on a set of defined assumptions and violations of these assumptions can have significant consequences on the reliability of the results. In selecting a specific model, the researcher is thus implicitly introducing a certain degree of model bias. For example, an assumption underlying most machine learning algorithms is that the data sample is drawn independently from the same identical distribution (Dundar, Krishnapuram, Bi and Rao, 2007). Different learning algorithms require a different sample complexity in order to sufficiently converge to an acceptable solution, and lower bounds for various algorithms have been established in the literature (Hanneke, 2016). This required sample size depends on a variety of factors such as model complexity, model-specific assumptions, the amount of (hyper)parameters in the model, and the signal-to-noise ratio within the data (Jordan and Mitchell, 2015). In summary, it is important when using machine learning to take into account that each machine learning model has a bias and learning algorithms whose bias best fits the data will perform better. Machine learning is more successful in some applications than in others because of differences in the underlying data and the nature of the task. It is therefore important to take into account the underlying assumptions of machine learning algorithms, understand the data, and recognize potential bias in the results.

Researchers should take into account the idiosyncrasies of the financial market environment when applying machine learning as it could impact the results significantly. Finance tends to provide an arduous environment for machine learning algorithms as various commonly-used assumptions are (mildly) rejected within the financial market environment. These elements are not necessarily an unavoidable obstacle for using machine learning in finance but are an important factor to keep into account. Financial markets are dynamic throughout time and are constantly subject to change. This implies, in statistical terms, that the data generating process is non-stationary as the joint probability distribution changes throughout time. This dynamic character is caused by three forces which are inherently present in the market. First one is a change in the market environment which causes relationships to structurally change temporarily or permanently (Ang and Timmermann, 2011). The second one consists of time series patterns such as trends and seasonality which cause relationships to follow a short-term trend or pattern in time. Finally there is the phenomenon in which the relationship disappears because it is arbitraged away or crowded out (Israel, Kelly and Moskowitz, 2020). Research has shown that inefficiencies and alpha returns tend to decrease after their publication which supports the argument that investors learn from mispricings and arbitrage them away (Mclean and Pontiff, 2015). This last argument is linked to the low signal-to-noise ratio of financial features which is negatively influenced by both the signal and noise component (Israel, Kelly and Moskowitz, 2020). The low signal is an inherent result of the market mechanism in which financial markets quickly incorporate new information and inefficiencies are crowded out by arbitrageurs. The low signal is thus not necessarily a fundamental feature of the data but exists by design as a result of the market mechanism. Black (1986) introduced the concept of noise as a form of anti-information where trading activity occurred independent of fundamental information. The high noise and randomness in financial data is a result of the real-time consensus mechanism of markets fueled by active short-term trading activity, new information entering markets, and human sentiment. Financial markets are not governed by predefined fixed rules as they are the aggregate net result of human interactions which make them subject to irrational and erratic movements. Markets are thus subject to human decision making which is inherently difficult to quantify and adjust for. Furthermore, markets are affected by a variety of exogenous and difficult-to-predict variables such as pandemics, natural disasters, political events, etc. There is an element of uncertainty, or unmeasurable risk, which makes markets rather difficult to predict. As a result of this uncertainty, finance data is generally characterized by fat-tailed distributions and high sigma events happen relatively frequently. All these elements make interpretability an important factor in financial models and the selected models should be appropriately scrutinized. Finally, a general issue is the limited data availability for many variables of interest in terms of the breadth of tradable securities (asset span) and time series length. For a specific financial variable, there is usually only one (limited) time series realization rather than a multitude of different realizations. As discussed by Israel, Kelly and Moskowitz (2020) finance is fundamentally a time series discipline and not providing the circumstances for big data analysis. Combined with the strong interconnectedness of financial markets which makes many features correlated, this contributes to the curse of dimensionality in finance. Given the aforementioned elements it is thus important to take into account non-stationarity, limited data availability, risk of overfit and detecting spurious relationships, curse of dimensionality, appropriate model selection and interpretability of the models when applying machine learning for financial research. We address these challenges in the rest of this section.

### 3.1 Non-Stationarity

For a long time, time series analysis has been a relatively small part of the machine learning literature as most methods have been developed for large cross-sectional data environments and point predictions. However, by borrowing mainly from advancements in sequence modeling, largely in the context of natural language processing, significant progress has been made in applications of machine learning for time series forecasting. Facebook's Prophet (Taylor and Letham, 2017) and Google's AutoML for time series (2020) are signs of increasing interest in the area. Although these models might not be optimal for the aforementioned peculiarities of finance, they do show an opening within the field of time series forecasting where machine learning could contribute greatly. Machine learning performs better for tasks which are static in nature as abruptly changing data distributions add an additional layer of complexity to the problem. Non-stationary data poses a challenge for machine learning as the majority of learning algorithms assume stationarity of the data generating process. However, adjustments can be made for non-stationarity and machine learning can be used to provide new ways of tackling non-stationary time series.

Conventional methods within econometrics dealing with non-stationarity often consist of transforming the time series via detrending, seasonality adjustments, and filtering. An important distinction is made whether the trend is deterministic or stochastic and different methodologies are applied based on this variation. Virili and Freisleben (2000) discuss different pre-processing techniques to handle non-stationarity before using the time series in neural networks. These may include taking single or compound power transforms, transformed differences, and wavelet transforms to decompose multi-frequency components (Lahmiri, 2014; Chandar, 2016). These techniques, however, transform the original time series and may remove valuable information in the process. As argued by de Prado (2018), there is a trade-off to be made between stationarity and "memory" within the data. Transformations, such as differencing, remove memory from the data and this removal of potential long-term dependencies may reduce rather than enhance the signal-to-noise ratio. De Prado (2018) proposes fractional differentiation for the purpose of financial machine learning to achieve stationarity but preserve historical information. However, transformation of data can sometimes act as a beneficial denoising method as argued by Balcerak and Schmelzer (2020). They show that by transforming financial time series into oscillations between regimes, considered labeled classes, one can leverage machine learning methods known to be successful in classification tasks. Although these approaches can be understood as emphasizing important information contained in the time series upfront, it is often difficult to justify the specific forms and dimensionality of the explicit transformations. This is where machine learning can be useful as it offers flexible data-driven models.

Bontempi, Ben Taieb, and Le Borgne (2013) conducted an empirical comparison between the most popular supervised learning techniques in time series forecasting and found that neural networks tend to perform better on several datasets. Machine learning approaches offer more flexibility in modelling the underlying process which can trigger regime shifts. The underlying process is determined implicitly in a data-driven manner. For example, dynamic support vector machines can learn different patterns for regime changes directly from the data (Cao and Gu, 2002). Shalizi, Jacobs, Klinkner, and Clauset (2011) proposed ensemble methods where the contribution of the different models can vary over time to take non-stationarity into account. Lahmiri (2014) combines wavelet transforms and neural networks to characterize the low-frequency long-term trends and high-frequency ruptures of the data separately. Similarly, Huang (2010) merges wavelet transforms and a recurrent self-organizing map (RSOM) algorithm to partition and store the temporal context of the feature space. Consequently, multiple kernel partial least square regressors are used as local models per partition to construct a final financial forecasting model. Giles, Lawrence and Tsoi (2001) have used recurrent neural networks to process financial non-stationary time series. Cheng, Sa-Ngasoongsong, Beyca, Le, Yang, Kong and Bukkapatnam (2013) provide an overview of the available toolbox for time series analysis in non-stationary environments.

One of the key innovations in non-stationary time-series modeling has been the introduction of computationally tractable recurrent network architectures that can model long-term dependencies, in the form of Long-Short-Term-Memory networks (LTSM) and Gated Recurrent Units (GRU). In RNNs, the recurrent connections add states to the network and allow them to learn broader sequential abstractions through memory. However, simple RNNs blend both long-term and short-term states through a single mechanism in which long-term states are to be gradually accumulated over time. The vanishing gradient problem places an effective limit to the range with which context can truly be accessed or learned, hence, time dependence in simple RNNs tends to be rather short-term. This is problematic in financial time series where the longer-term underlying state or regime of the economy and cyclicality can change the dynamics and direction of interactions. LSTMs and GRUs circumvent these problems via gated units that explicitly control the update mechanism. With a suitable number of recurrent units and network depth, LSTMs and GRUs can model rich and flexible temporal dependence and dynamic state-dependent behavior. Zhao (2017) uses temporal proximity weights and an LSTM network to predict stock price trends and shows that the weighted LSTM outperforms competing support vector machines, random forest, and regular RNN models. Sirignano and Cont (2018) use a deep LSTM architecture to predict the direction of the next price move in a high-frequency setting, using the entire state of the NASDAQ limit order book. The authors report that universality and relative stationarity is achieved without the input preprocessing and segmentation that is required in traditional models. Some studies explicitly exploit the fact that LSTMs embed a summary of the variable history as a state. Chen, Pelger, and Zhu (2019) utilize an LSTM to derive a small set of economic state processes from more than 100 initial macro-economic variables. These state processes are then fed into a wider network architecture to set the dynamic context of interactions of company-specific features in the cross-sectional asset pricing problem. Temporal Convolutional Networks (TCN) with dilation (Lea 2016, Deng 2019) have also surfaced as a viable candidate for dynamic temporal dependence modeling. These architectures can be stacked to express rich multi-scale dependencies and also be incorporated into wider network architectures.

One important group of models that have been used to handle dynamic time series have been regime switching models. These can be cast into two groups, namely, threshold models and hidden Markov models (Piger, 2007). Both of these model families impose a rigid structure on the time-dependent data generating process. Threshold models assume that the regime shifts occur when a particular quantity passes an unobserved threshold, and hidden Markov models assume that regime shifts are governed by an unobserved Markov process. Markov-shifting models have been a popular regime switching framework for financial and economic time series. However, in these models the choice of the number of regimes is either arbitrary or loosely based on domain knowledge, the transition probability matrix estimations can be unstable if regime transitions are infrequent, and large state-dependent empirical covariance matrices have to be estimated. Furthermore, the nature of the regime changes of the underlying data generating process may exhibit non-Markovian path dependency. These disadvantages are in contrast to the previously mentioned LSTM and GRU type models whose flexible internal states enable the expression of complex long-term dependencies. Recently, some researchers have overlayed Markov switching on top of RNNs and other network architectures to explicitly model the temporally varying dynamics of financial variables (Ilhan, Karahmetoglu, Balaban and Kozat, 2020; Bildirici and Esrin, 2016), which could be an approach when there is some clarity over the structure of the regime delineations.

A certain type of regime shift in finance is particularly difficult to handle, namely, temporary regime shifts which occur as a consequence of a fat-tailed crisis event. Crisis events tend to have considerable consequences as they temporarily dislocate markets and fundamentally alter relationships between financial covariates. Fat-tail events often require special treatment and should be considered as a modelling challenge different from structural regime shifts as they are particularly difficult to predict. This is caused by the rarity and the severity of the distributional deviations as well as the uniqueness of each individual crisis. It should be noted that most models do not properly capture heavy-tailed events and tend to inaccurately take them into account (Wang, Varul and Eliassi-Rad, 2019). Taleb (2020) formalizes the issues of handling heavy-tailed distributions, by emphasizing the underlying assumptions that most conventional estimation techniques take and which do not hold for some parameters of fat-tail distributions. Modern machine learning methods can offer a solution as they do not seek to estimate parameters that characterize the fat-tail distributions but rather mimic the underlying data generating process. One difficulty is that there is only little data about these fat-tail events to learn from. This is a limited data problem which could be mitigated by using synthetic data (e.g. generated through generative adversarial networks), augmenting the observation space with spatial-temporal cohort analysis where applicable, transfer learning, and class imbalance resolution mechanisms.

### 3.2 Limited Data

A general issue within finance is the limited data availability in terms of the breadth of tradable securities (asset span) and the time series length. For some important financial variables, such as policy interest rates, there is only one limited time series realisation available. There are only a limited number of traded assets available in the market and new objects (assets) cannot be easily added or created by the researcher. Furthermore, newly available big data sets tend to provide limited time series which makes it difficult to make statistically significant linkages with the financial-economic variables of interest. As argued by Kelly, Israel and Moskowitz (2019) finance is fundamentally not a big data environment compared to other fields where machine learning successes were achieved. Machine learning algorithms require sufficient data to provide unbiased results and this required sample size tends to be large for some algorithms. Some of the most popular learning algorithms, such as deep neural networks, require large amounts of data. Van der Ploeg, Austin and Steyerberg (2014) compare the data requirements of different machine learning methods and find that popular algorithms need over ten times the number of data points per feature in order to provide stable results relative to traditional statistical techniques. Cerquiera, Torgo and Soaras (2019) observe that machine learning outperforms traditional statistical methods when the sample size grows and this sample size is of significant importance when choosing between forecasting models. In this regard, the researcher needs to select the appropriate algorithms given the data in the specific financial problem. For example, analyzing high frequency order book data will allow for a broader variety of machine learning algorithms to choose from compared to analyzing monthly series of macro-financial data. Some machine learning algorithms are more suitable for "small" data environments and one could manage this further by controlling for the number of features and hyperparameters.

First and foremost, limited data availability increases the risk of machine learning algorithms to overfit as they try to learn the underlying structure of the data. The problem of overfitting and the low signal-to-noise ratio gets thus worse in the face of limited time series. Cerquiera, Torgo and Soaras (2019) argue that machine learning methods generate biased time series forecasts when the sample size is limited as they show tendencies to pick up spurious relationships. It should be noted that limited data availability poses a challenge to some popular methods to manage overfit such as cross-validation and that appropriate adjustments should be considered. One important way to manage the required data size is to perform dimensionality reduction. However, limited data makes feature selection itself more difficult to perform because the data is sparse within the dimension of each feature. Dimensionality reduction is all the more necessary in the context of small data since data sparsity is more problematic when the considered volume in the space increases. In the face of limited data, it is thus important to manage overfit and dimensionality reduction which are both separately discussed later in this section.

The small data environment causes several other problems such as class imbalance, distributional asymmetry, and limited information availability on certain infrequent events (Choi, 2019). Crisis events, which tend to have a significant lasting impact on markets, can cause a significant class imbalance in financial time series. A substantial portion of the underlying distribution is never realised because we only ever observe a few realisations of the data generating process. Hence, the infrequent occurrence of fat-tail events creates both data sparsity and class imbalance (Choi, 2019). In the machine learning literature, there exists a number of techniques to deal with data imbalance such as undersampling, oversampling, weighted classes and weighted loss functions. Class weights could also be learned like has been done by Liao, Shih, Chen and Hsu (2014) who have used support vector machines as a pre-processing method to learn the best class weight.

A promising approach to tackle limited data availability is (Bayesian) transfer learning where one uses prior information in the learning process. One could use theoretical considerations, findings in the literature and domain expertise to select a task where one would expect similar dynamics in the underlying data generating process, and use transfer learning to learn another data generating process where less data is available. Transfer learning is thus particularly promising for the purpose of reducing the difficulties with small data. Technically, it consists of starting the training procedure with a model that is pre-trained on a data generating process that shares commonalities with the data of interest. This way, the model would benefit from patterns learnt in a setting that has better data availability and which is assumed to be somewhat similar. The model is updated with the small data which reinforces the patterns that match and reduces the confidence in the patterns that cannot be seen in the small data of interest. For example, Jeong and Kim (2019) improved trading strategies on small stock indices where they leveraged on transfer learning from a large universe of stock indices. He, Pang and Si (2019) demonstrated the benefits of transfer learning on baseline methods by utilising features extracted from different data sources to make predictions on several financial time series.

One way in which researchers could potentially alleviate the limited data issue is to generate synthetic data. Synthetic data is increasingly used to train algorithms and machine learning itself, in its ability to learn the underlying structure and properties of the data, is a suitable candidate to generate synthetic data. Synthetic data can also provide artificial variability to the data which would make algorithms more robust to overfitting. The application of synthetic data is more useful in some areas of financial research than in others. It would be, for example, mainly advantageous for research topics such as portfolio management, market microstructure, and risk analysis. Synthetic data generation essentially consists in finding a model for the underlying data generating process, and generating new data according to this model. de Prado (2020) provides an overview of synthetic data generation methods for time series in which the two main approaches could be divided between resampling and Monte Carlo methods. Especially recent advancements in adversarial neural networks provide opportunities for synthetic data generation. For example, TimeGAN introduced by Yoon, Jarrett and van der Schaar (2019), enables the generation of data according to a best-fit neural network approximator of the real data while preserving the temporal dynamics of the process.

Finally, it should be noted that certain traditional statistical measures cannot not be estimated in an unbiased way with limited data availability. Small data creates undesirable bias in the estimation of some parameters, as the assumptions taken by a lot of conventional estimation techniques (e.g. central limit theorem) do not hold anymore. Bayesian and transfer learning can help resolve this by introducing a competing bias, based respectively on domain knowledge and data from a similar problem, which would compensate for the unwanted bias of small data.

Machine learning methods could thus be used to provide a solution for the difficulty of unbiased estimation of conventional statistical measures in the face of limited data. One example in finance is the estimation of the variance-covariance matrix which is used for various (empirical) financial applications from portfolio construction to factor analysis. de Prado (2016) calls this the Markowitz curse which is the instability of the inverse covariance matrix when assets are correlated. This difficulty of estimating expected returns, where there is a large sampling error using limited time series, was already noted by Merton (1980). de Prado (2016) applies machine learning techniques to compute the correlation structure of the assets.

### 3.3 Overfit

While machine learning allows to capture more flexible relationships, this is also its weakness as it could pick up spurious patterns. The low signal-to-noise environment in financial markets naturally poses difficulties as machine learning algorithms tend to be computationally capable of identifying patterns whether they are spurious or not (de Prado, 2020). The elegant linear methodologies in econometrics are generally less susceptible to find these random patterns. As a result, machine learning puts a substantially larger emphasis on out-of-sample generalizability than conventional econometrics, and generally requires more attention to overfitting (Athey and Imbens, 2019). The financial market environment makes handling spurious relationships and robustness more important than in other machine learning applications. As discussed by Arnott, Harvey and Markowitz (2018), in the age of machine learning financial researchers require a more robust empirical testing protocol to avoid overfit. Validation in empirical research in econometrics tends to happen in-sample while validation in machine learning happens out-of-sample (de Prado, 2019). This out-of-sample validation methodology in machine learning in itself reduces the potential overfit of a model and makes it less sensitive to the sample. Common machine learning practice is to work with training and test data samples in order to examine how well the model generalizes. de Prado (2020) argues that overfitting can happen in both the test set and the training set. The former could be alleviated via ensemble methods, regularization methods and cross-validation while the latter could be alleviated through cross-validation. This sensitivity to find spurious relationships also brings us to the importance of methods for model validation, feature selection, model choice, and interpretability which are discussed later in this section.

Cross-validation is a methodology in machine learning involving resampling of the data sample in various subsamples. However, these cross-validation techniques have been generally developed for big cross-sectional data environments. The limited data availability in finance limits the possibilities of using prominent techniques against overfitting such as K-fold cross-validation (Arnott, Harvey and Markowitz, 2018). Financial researchers could already count themselves lucky if they have gathered enough data to properly train the machine learning model, let alone splitting the data set into a number of subsamples. Vabalas, Goven, Poliakoff and Casson (2019) confirm that K-fold cross validation produces biased results and overfitting in limited data environments. They suggest the use of nested cross-validation, where model selection and hyperparameter tuning is performed at the same time, which would produce unbiased results regardless of sample size. This illustrates the importance of robust hyperparameter optimization and to incorporate this in the cross-validation exercise. Moreover, one could argue that traditional cross-validation methods are rather inappropriate for time series data where there is a temporal order. Cerqueira, Togo and Mozetic (2019) show that blocked cross-validation works well in the case of stationary time series and Rep-Holdout cross-validation performs adequately in the case of non-stationarity. It is thus important to adjust the cross-validation methodology to make it suitable for financial time series. More robust research on this topic of cross validation in financial-economic time series would be required. It is important to note that in machine learning one needs to find a right balance between model complexity and model performance. Hyperparameter selection has an effect on model complexity and could be handled via cross-validation.

Ensemble methods allow to combine a set of different learning algorithms or the same learning algorithm with different (hyper)parameters and features. Ensemble techniques could be especially useful for prediction in finance where there is model uncertainty and risk of overfit. Machine learning models suffer from three errors in the form of bias, variance and noise, and ensemble methods help to tackle the former two (de Prado, 2018). It should be noted that model selection via ensemble would be mainly applicable for the purpose of prediction tasks instead of empirical testing where one wants to test a particular model. For the aforementioned reasons we observe outperformance of ensemble methods in several prediction competitions and random forests are considered as one of the most popular machine learning methods (Crone, Hibon and Nikolopoulos, 2011). There are several ensemble methods such as bagging, boosting, stacking and Bayesian model averaging. Bagging involves equal weights, boosting uses weights based on mis-classification, stacking tries to learn the optimal weights, and Bayesian model averaging involves weights based on posterior probabilities of each model. Adjustments have to be made to the conventional ensemble methodology given the non-iid time series nature of financial data. As argued with cross-validation, similar adjustments could be made in the ensemble procedure. For example, de Prado (2018) argues that bagging is preferred to boosting in this case, suggests to apply sequential bootstrapping instead of bagging, and asserts to not apply reshuffling.

Regularization methods can mitigate overfitting of machine learning techniques as they manage the complexity of the model via penalization of the loss function. Appropriate feature selection, which is discussed in the next subsection, is one of the key elements for avoiding overfit in machine learning models and regularization could be seen as a form of implicit feature selection. Regularization is a method to "regularize" the effect of the selected features within the model and thus simplifies the model to tackle the bias-variance trade-off. Popular methods for regularization of the loss function include lasso (L1), ridge (L2), and elastic net (combination of L1 and L2) which provide a statistical approach to weigh down the coefficients of less important features. Moreover, there are specific regularization methods for certain algorithms such as dropout in neural networks. Within conventional econometrics a similar principle to regularization is used via information criteria, however, the difference with regularization within machine learning is the focus on out-of-sample predictive performance (Athey and Imbens, 2019). Penalized regression algorithms are preferred in the case of feature sparsity which makes it suitable for a variety of financial applications such as asset pricing (Athey and Imbens, 2019). This is confirmed by Chinco, Clark-Joseph and Ye (2017) who show that regularized regression outperforms in modelling cross-sectional returns because of the importance of short-term and sparse features in asset prices. Evidence also indicates that regularized regression methods outperform traditional regression techniques in low signal-to-noise environments (Athey and Imbens, 2019). Finally, it should be noted that regularization adds another parameter (i.e. regularization parameter $\lambda$) to the model and thus requires appropriate tuning. Tuning of this parameter is generally performed using cross-validation or gradient descent approaches (e.g. Feng and Simon (2018)).

Although flexibility in the specification is a vaunted advantage of machine learning algorithms, the imposition of some structure based on economic theory can aid both prediction and inference in low signal-to-noise environments. Israel, Kelly and Moskowitz (2020) argue that combining theory with machine learning could mitigate the issue of overfitting and spurious relationships. One could start with economic structure and apply machine learning in aspects of the model where it fails or where the current state of theory is insufficient. Since both the set of relevant asset pricing factors and model choices are very large, economic theory can guide architectural choices such as the incorporation of temporal recurrence, spatial convolutions, feature partitions, and conditioning covariates. Gu, Kelly and Xiu (2019) show improved predictions by embedding theoretical considerations such as no-arbitrage and the linkage between return and risk in equilibrium in an autoencoder model. More specifically, they use a conditional autoencoder structure where factor loadings are nonlinear functions of asset characteristics and the factors themselves are portfolios of individual stock returns. Chen, Pelger and Zhu (2020) pose the cross-sectional asset pricing problem as an estimation of the stochastic discount factor with no-arbitrage constraints. These economic constraints serve the purpose of regularization in separating the risk premium signal from noise. The macroeconomic time-series covariates are separately summarized from the latent states of an LSTM network which are then connected to a feed-forward network alongside firm-specific cross-sectional attributes. It should be noted that in both studies, the structural formulations and economic constraints (e.g. no-arbitrage, no-intercept, and factorization) not only serve as regularizations but also hypothesize economic mechanisms that can be tested.

Finally, empirical researchers should make adjustments to their research protocol given the higher risk of overfit using algorithmic methods. Several methodologies have been proposed to reduce backtest overfitting in asset pricing research. Traditionally, assessments are made in-sample which encourages p-hacking through multiple testing. This phenomenon of p-hacking has been a difficult problem faced by the financial research community (Harvey, 2017). Arnott, Harvey and Markowitz (2018) discuss an adjusted protocol to perform empirical research in asset pricing with machine learning. Bailey, Borwein, de Prado and Zhu (2015) propose combinatorially symmetric cross-validation to examine the probability of sample overfitting. Bailey and de Prado (2014) propose the deflated Sharpe ratio which adjusts for selection bias as a result of multiple testing and non-gaussian return distributions. It is important to acknowledge that adjustments to the empirical research process have to be made and that a greater focus on out-of-sample methodologies would be required. Applying machine learning methodologies and the philosophy of out-of-sample generalizability could help combat p-hacking, and thus in-sample overfitting, in financial research. Machine learning could thus help to bring positive change to how empirical research is done and judged by financial economists with a broader focus on out-of-sample generalizability.

### 3.4 Curse of Dimensionality

The curse of dimensionality relates to the exponential increase in the required amount of data to sufficiently train a model when the dimensionality increases. When increasing the amount of features, the data becomes more sparse across the different dimensions. It is strongly present in finance given the high dimensionality in terms of potential explanatory features and relatively limited time series data for these variables. During the past decades, the literature has garnered a wide variety of variables to explain financial relationships. Especially within asset pricing there is a large variety of factors which have been published. Researchers have developed a factor zoo (Harvey and Liu, 2020) of hundreds of factors systematically explaining asset returns. To make things even more statistically challenging, many financial variables tend to strongly correlate with each other. For example, asset returns tend to be highly correlated and can be mainly explained with only a handful of features. Israel, Kelly and Moskowitz (2020) demonstrate that only three principal components explain about 80% of the variation in Fama-French portfolios. This observation that most features are actually redundant in a multivariate setting introduces the concept of sparsity which is a direct result of competitive markets and is related to the argument on the low signal-to-noise ratio of financial variables. Traditional regressions methods cannot properly handle high dimensional data or data with strongly correlated variables. More specifically, it leads to overfitting, lower accuracy, spurious collinearity of features, and inappropriate measurement of the relevant parameters (Fan, Lv and Qi, 2011). Some machine learning algorithms are more sensitive to the curse of dimensionality than others and preprocessing of the data with respect to the features would be necessary in order to use them. Clustering algorithms are one example of methods which have difficulties with the dimensionality curse and require appropriate a priori feature selection. The curse of dimensionality also affects machine learning algorithms because successful learning requires sufficient amounts of data to cover the space in which the model has to hold (Verleysen and François, 2005). It is thus important to mitigate this dimensionality and apply methodologies that are suitable for sparse feature settings. To handle the curse of dimensionality, researchers could use dimensionality reduction techniques during processing or techniques to control for relevant features such as penalization. Moreover, small data samples themselves are adverse environments for many commonly used dimensionality reduction techniques which is important to take into account.

The reduction of dimensions in the processing stage has several advantages such as reduced tendency of models to overfit the data, enhanced model parameter estimation as only relevant features are being considered, and improved model interpretability as redundant variables are removed. As argued by Athey and Imbens (2019) econometricians tend to hand-select the relevant variables while data-driven feature selection can provide improvements in the face of (approximate) sparsity. Appropriate dimensionality reduction should be considered as a key part of building financial models and data-driven methods could be a powerful tool in this regard. Dimensionality reduction could be achieved through feature selection or feature extraction. Feature selection is concerned with selecting the most relevant features while feature extraction deals with transforming features into a lower dimensionality representation. Feature extraction has a relative advantage that dimensionality reduction can be achieved without much information loss of the original feature space but a relative disadvantage in terms of difficult interpretability of the new features (Khalid, Khalil and Nasreen, 2014).

The literature provides a variety of feature selection methods which could be broadly divided into filters, wrappers and embedded methods (Cai, Luo, Wang, and Yang, 2018). However, traditional feature selection techniques tend to be unstable in small data environments (Dernoncourt, Hanczar, and Zucker, 2014). Furthermore, as argued by Fan and Lv (2010), collinearity provides difficulties in feature selection as this can be spurious in high dimensional geometry which could result in overfitting and inappropriate variable selection. As argued by Dernoncourt, Hanczar, and Zucker (2014), this could be potentially solved by using a priori knowledge and unsupervised machine learning methods to filter out the most irrelevant variables. More research would be needed in this area of feature selection in small data environments and this shows that traditional techniques used in machine learning cannot be easily converted to finance. Machine learning models themselves, coupled with feature importance analysis, can be used as a viable filtering method for feature selection. The researcher can start with a wide variety of potentially relevant features together with an easy to tune machine learning algorithm which could be used to rank features using relative feature importance. This preliminary model can be trained to generate relative feature importance values before advancing to more complex machine learning models. At this stage, an analysis of the presence of nonlinearities and cross-effects could guide the architectural choices for the final model. Furthermore, feature selection through ensemble techniques, where several methodologies are combined or applied on different sample subsets, could be used to generate a more robust feature selection (Chandrashekar and Sahin, 2014).

Factor analysis (FA), principal component analysis (PCA), and linear discriminant analysis (LDA) have been traditionally used in finance to extract features. Machine learning measures provide more flexible alternatives to these methods which are fundamentally based on linear matrix factorization. In this regard, manifold learning methods and autoencoders can be used to construct lower dimensional representations in a non-linear fashion. It should be noted that one could use nonlinear variants of PCA such as Kernel PCA. Manifold learning algorithms such as Isomap, Locally Linear Embedding (LLE), and Maximum Variance Unfolding (MVU) transform the data set into a lower dimensional one while trying to maintain the initial geometry of the data (van der Maaten, Postma and van den Herik, 2009). On the other hand, autoencoders are an artificial neural network that learns an efficient encoding and maps a high-dimensional input space into a low dimensional intermediate space. Van der Maaten, Postma and van den Herik (2009) compare dimensionality reduction techniques and find that autoencoders perform well among their peers. An disadvantage of autoencoders is that they require sufficient amounts of data as small data samples cause high variance in the gradients (Liu, Wei, Zhang and Yang, 2017). To handle small data samples, one could apply regularized autoencoders which force sparsity via regularization of the nodes in the network. For example, Bao (2017) uses stacked autoencoders as a mechanism to de-noise and extract high-level hierarchical features ahead of a final LTSM model to forecast prices of market indices. Similarly, the internal feature maps of temporally convolutional networks (TCN) and the hidden states of LSTMs can also be considered lower dimensional representations of multivariate temporal dependencies. Borovykh (2018) uses a TCN-dilation architecture to forecast the stock market conditional on macro-financial control variables, and Chen (2020) applies a LSTM to find a small set of temporally dependent economic state processes from more than 100 initial variables. Besides providing greater functional flexibility, dimensionality reduction can provide additional advantages such as interpretability. For example, Bryzgalova, Pelger and Zhu (2019) argue for the application of decision tree pruning to select relevant asset pricing portfolios. They argue that it offers similar insights to PCA while offering more interpretable results for the purpose of asset pricing.

As aforementioned, regularization is a method to "regularize" the effect of the selected features within the model and thus implicitly selects the most appropriate features. These regularization methods force irrelevant features to zero within the optimization procedure which effectively leads to a sparse model. It should be noted that ridge does not properly regularize the coefficients of less appropriate features to zero (Melkumova and Shatskikh, 2017). In the case of feature selection in sparse environments, lasso and elastic net are argued to be the preferred method of regularization (Muthukrishnan and Rohini, 2016). For this reason have penalized regression methods been used to handle high-dimensional feature sets while keeping the advantages of conventional regression methods such as interpretability (Tibshirani, 2014).

### 3.5 Model Selection

Most traditional models in financial economics are based on structured hypotheses where a researcher proposes a hypothesis about the relationship between different financial variables under a specific functional form and distributional assumptions. The functional forms typically involve linearity, additivity, separability, and effect monotonicity. Possible transformations such as logarithmic, exponential, differencing, integration, thresholding, and lags are conducted explicitly. While the assumptions and functional restrictions improve model interpretability, parameter estimation, and the ability to apply statistical tests to specific interactions, they also carry considerable model misspecification risk (Jung, Patnam and Ter-Martirosyan 2018). In conventional econometrics there is a focus on parameter estimation of the pre-specified model with much less emphasis on systematic validation and comparison across different modeling frameworks (Athey and Imbens 2018). Many modern machine learning algorithms are non-parametric by nature, and are geared towards a data-driven and goal-oriented approach that prioritizes prediction accuracy. However as there is a large variety of different model families and setups in the form of model architecture and hyperparameters, one requires a disciplined framework for model selection.

The first step in selecting an appropriate machine learning model is to categorize the problem at hand. The researcher determines whether the input data is labeled with the intended prediction target for a supervised learning problem, unlabeled with the intent to uncover intrinsic structure in the data for an unsupervised problem, or mixed between labeled and unlabeled data for a semi-supervised problem. On the output side, one needs to identify whether it is most appropriate to predict real values as a regression problem or categorical values, in the form of binary or multi-class variables, for a classification problem. The research focus can also be on dimension reduction, anomaly detection or estimating the generator of an underlying data distribution. Sometimes it is most effective to pose the problem at hand as learning an optimal policy in a sequential decision-making process, with feedback from the environment in a reinforcement learning setting. It is important to note that model selection in machine learning is heavily dependent on the application at hand. As will be discussed in the last section of this paper, machine learning could be used for different parts of the research process such as empirical testing, causality analysis, and prediction. Different methodologies and algorithms have been developed based on the specific task, each problem class entails a subgroup of appropriate modeling frameworks. For example, specific machine learning algorithms have been developed for causality analysis which are different from algorithms used in prediction. Applying machine learning is not one-size-fits-all and requires a more task-specific approach of researchers. To limit the scope of discussion to key elements of model selection typically encountered in financial research, we implicitly focus on model selection in a supervised regression setting.

In line with the "no free lunch" theorem (Wolpert and Macready 1996), no machine learning algorithm is superior to others in all scenarios. Hence, it is important to develop a robust pipeline of comparative candidate models that are appropriate for the problem at hand. Exploratory data analysis should be performed based on characterizing the univariate distributions, multivariate dependencies, outliers, and preliminary fitting of simple models. This exploratory data analysis provides the researcher with a sense of feature importance, spatial and temporal dependencies, the necessity of data normalization or winsorization, and possibilities for intermediate dimension reduction. It also helps to assess the appropriateness of certain distributional assumptions embedded in machine learning algorithms (e.g. spherical variance for K-means) which enables the researcher to expand or limit the model search space. In particular, for financial time series prediction problems one is often confronted with a) strong temporal dependencies and feedback mechanisms across time, b) integrated and sequence dependent raw level data, c) heterogeneous differencing orders across inputs, d) lagged dependence in difference-transformed data, e) embedding of broader market regime information in long-term dependencies which set the context for other interactions, and f) multitude of different time-scales of temporal dependencies. Alongside vector auto-regressive models, feed-forward networks, and tree-ensemble models with explicit lag-terms included in the input vector, one should also consider computationally tractable recurrent architectures such as Long-Short-Term-Memory (LSTM) networks and Gated Rectified Units (GRU) as candidates when dealing with longitudinal prediction problems. Similarly, one can also consider Temporal Convolutional Networks (TCN) with dilation (Lea 2016, Deng 2019) to incorporate long-term dependencies. In this regard, Gu, Kelly and Xu (2020) compare different standard machine learning models to the cross-asset pricing problem.

Complex models with many parameters can often overfit the peculiarities of the sample in ways that fail to generalize. Hence, a model selection framework should focus on generalization performance on curated hold-out datasets separate from those used in training. This entails a two-stage process as there is model selection within a given model family in the form of hyperparameter optimization as well as the outer evaluation across hyperparameter-optimized model families. For instance, within Gradient Boosting Tree models, one has to determine the optimal number of estimators, learning rate, and maximum depth. Within an LSTM model one needs to select the number of recurrent units, network depth, and dropout rates. In financial time-series applications, due to both data scarcity and temporal dependence, a three-way hold-out method is typically used where the data is contiguously split into training, validation, test sets. The validation set is used for hyperparameter selection whereas the test set is used for final model evaluation. Because of temporal dependence, in order to prevent influence leakage, it may be prudent to have embargo periods that separate the three time segments. In cases where temporal dependence is less important, for instance when features and/or prediction targets are cross-sectionally normalized, a full-fledged M-by-N nested cross-validation can be conducted. Here both hyperparameter tuning and evaluation are conducted by K-fold cross-validations, where the inner loop is responsible for hyperparameter selection and outer loop responsible for estimating the generalization accuracy (Varma and Simon, 2006). In K-fold cross-validation applications it is important to stratify the samples such that the folds have approximately equal distributions in terms of the target variable, as financial variables can have varying distributional properties over time. In general, simple models, such as linear regression models, tend to have high bias from underfitting whereas more complex models, such as neural networks, tend to have low bias but high variance from sensitivity to the training data. Data availability becomes an important decision factor as complex models are less likely to be effective when data is limited. Having a rigorous validation and testing framework enables the researcher to strike a balance in model selection in this bias-variance and underfit-overfit trade-off. Beyond generalization performance, there may be other aspects that are important in model selection such as interpretability and regime stability.

### 3.6 Interpretability

In machine learning there is a common dilemma between prediction accuracy and model complexity (Breiman, 2001). The potential accuracy improvements afforded by complex, high-dimensional, and nonlinear machine learning models often comes at the cost of less interpretable interactions. This inscrutability can lead to a rejection of model adoption by researchers and cast doubts about model generalizability. This could be especially the case in settings, such as financial markets, with limited observational data availability and relatively frequent regime changes. Machine learning algorithms, such as deep neural network architectures, offer the flexibility necessary to tackle the complex and temporally dependent interactions between variables encountered in many financial market problems. In particular, recurrent and convolutional networks have enabled the embedding of the temporal and spatial dependencies between market variables in tractable ways. However, limited data availability coupled with a high number of effective parameters and compound nonlinear transformations raise concerns about overfitting to particularities of the noise within the sample. Hence, understanding what drives the predictions of more complex machine learning models and the ability to examine variable importance is a necessity for the purpose of financial research. This is not a mere convenience or bridge to traditional methodologies but an essential path to pursuing robustness, reliability, and replicability in the financial application of machine learning models. Embracing model interpretability as both an ongoing model-building framework as well as a suite of diagnostic tools enables researchers to complement model predictions with domain knowledge. This is particularly important in the face of low signal-to-noise ratios, non-stationarity, and the presence of outliers.

Capturing complex high dimensional relationships into more concise and understandable results has value in itself, even when the method may be a black (or grey) box. With recent developments in interpretable machine learning, many tools and methods have been proposed to improve model interpretability along both model-specific versus model-agnostic, and local versus global vantage points. As argued by de Prado (2020), whether machine learning is a black box depends on the person who is using it and not on machine learning algorithms themselves, which are powerful tools open to understanding, to explicitly provide it. We acknowledge that some algorithms may be less interpretable than others and that the interactions may be more difficult to understand than simple linear ones. But if one understands the underlying mechanics of the learning algorithms and uses the tools at one's disposal to interpret them, one could have a broad understanding of what the algorithms are capturing. In this regard, machine learning can help to get a better understanding of the complex, non-linear and multivariate relationships present in financial markets.

When interpretability is achieved by restricting the complexity of the model itself, this is referred to as intrinsic interpretability (Molnar 2019). Most traditional econometric models such as linear factor models in asset pricing, vector autoregressive models for longitudinal predictions, GARCH variants in volatility modeling, linear state-space models, and Markov switching models are intrinsically interpretable. As previously discussed, traditional econometric model specifications are primarily motivated by amenability to statistical inference. The underlying data generating process is characterized as a probability model, which is proven to be theoretically sound under certain assumptions, allows to statistically infer some characteristics and easy interpretation. They are usually endowed with conditional linearity, additivity, separability, and effect monotonicity. Transformations such as logarithmic, exponential or differencing are conducted explicitly. The interpretability of these models is such that a simple generative process can be prescribed in terms of the relationship between innovations of covariates and for the ease of testing the statistical significance of underlying hypotheses via parameter p-values. However, as pointed out in de Prado (2020), p-value based significance testing relies on strong distributional assumptions, can generate misleading results in the presence of collinearity, and tests parameter significance conditional on the model being correctly specified. The easy interpretability in conventional econometrics can thus be a poisoned chalice leading to wrong interpretations when the standard assumptions are violated.

In recent years, researchers have developed approaches to incorporate interpretability in modern machine learning algorithms by structurally enforcing some of the interpretability constraints mentioned before. Vaughan (2018) proposes explainable neural networks (XNN) by limiting the connections between nodes such that the learned network model is a modified additive index model. Explainable boosting machines (EBM) introduced by Nori (2019) is a generalized additive model that incorporates bagging and gradient boosting to have accuracy comparable to random forest and (unconstrained) boosted tree models.

Post-hoc interpretability, on the other hand, refers to techniques and tools that can be applied to the models after training, usually to summarize the impact and importance of different input features. Some post-hoc interpretability techniques are specific to the model. In case of tree-ensemble models, such as random forests and gradient boosted trees, feature importance can be measured as mean decrease impurity. Alternatively, since random forests and the stochastic variant of gradient boosting use bagging, the feature importance can be computed on permuted out-of-bag (OOB) samples based on the mean decrease in prediction accuracy. In the case of neural network models, DeepLIFT (Shrikumar 2019) backpropagates the contributions of all neurons in the network to every feature of the input and compares the activation of each neuron to its reference activation while assigning contribution scores according to the difference. Horel (2020) approaches the neural network interpretability problem from the perspective of statistical significance testing by constructing a gradient-based test statistic of estimators and deriving asymptotic distributions. Combining deep neural networks with post-hoc interpretability methods can reveal significant non-linearities and cross-effects that cannot be assessed in traditional asset pricing models.

While model-specific interpretations are important, machine learning research is often conducted across multiple competing models. Hence, there is a need to elucidate what drives prediction mechanisms in a model agnostic way. Partial dependence plots (PDP) show the marginal effect that a set of features have on the predicted outcome by averaging out (in Monte Carlo sense) the effects of all other input variables over sample. They are particularly useful in visualizing linearity, monotonicity, domains of positive or negative responses, and two-way interactions. Individual conditional expectation (ICE) is a good complement to PDPs as it demonstrates how a model behaves per observation. It provides local information that complements the global information of PDP. Accumulated Local Effects (ALE) demonstrate the overall behavior of the predicted outcome with respect to an input variable. ACE plots circumvent the feature collinearity problem of PDP by averaging the changes in predictions over the conditional distribution and accumulate them over a local neighborhood. ALEs can be used in conjunction with PDPs when input collinearity or strong interactions are suspected. Permutation feature importance is another simple and intuitive model agnostic measure where the increase in prediction error of a model is computed after the values of a feature have been permuted. Since the permutation itself adds randomness, it is best practice for the permutations to be sufficiently repeated and importance measures averaged.

Local Interpretable Model-agnostic Explanations (LIME) is a local surrogate methodology where first, an auxiliary training data set is generated by perturbing the data around the instance of interest and computing the predictions from the target model. A model of choice from the intrinsically interpretable class (e.g. linear model) is then trained and local interpretability is inherited from this simplified representation. Finally, a unified local-global framework that has garnered significant empirical support is Shapley values. Conceptually, Shapley values characterize the fair attribution of a player's importance in a cooperative game. Computationally, it is the difference in the prediction value with and without a given feature averaged over all possible feature coalitions. Shapley values are appealing as it currently is the only attribution method that satisfies theoretical properties of local accuracy, missingness, consistency, and additivity (Lundberg, 2017). Lundberg (2017) proposes using Shapley additive explanation values (SHAP) as a unified framework of feature importance, where SHAP values are the Shapley values of a conditional expectation function of the original model. SHAP attributes to each feature the change in the expected model prediction when conditioning on that feature, explaining how to get from the base prediction to the current output, in both a locally and globally consistent manner. While Shapley values have gained significant traction as a post-hoc model-agnostic interpretability tool there remain challenges. Sundarajan (2020) and Kumar (2020) challenge the choice of using conditional expectations in its operationalization as it can lead to counter-intuitive outcomes. Ma (2020) analyzes the predictive and causal implications of Shapley values and demonstrates that they do not result in the most parsimonious and predictively optimal model in the general case.

As pointed out in Molnar (2020a, 2020b), there are many challenges to interpretability research and practice. Accounting for poor model generalizability and prediction uncertainty, feature dependence, and multiple comparisons are among those challenges. In addition, one must be mindful of the nuanced differences between perturbation feature importance, conditional feature importance, and relative feature importance and utilize the concepts most applicable to the questions being asked.

## 4. Applications of Machine Learning in Financial Research

Machine learning poses various opportunities for different parts of the research process including empirical discovery, estimation, empirical testing, causality analysis, and prediction. In this section we discuss how machine learning could be used in the various parts of the scientific process in finance.

### 4.1 Empirical Discovery

Machine learning is an ideal candidate for empirical discovery because of its ability to identify complex patterns and break down high-dimensional data into low-dimensional components. Machine learning algorithms are especially convenient in pattern detection with minimal supervision, and a large literature has been developed on unsupervised machine learning methods. Instead of using predefined mathematical simplifications to describe the reality of financial markets, one could use machine learning algorithms to discover empirical relationships from the data. In this approach researchers could look for variables or relationships in a complex system using machine learning algorithms, try to examine how these findings would fit in a generalizable financial-economic theory, and finally empirically test this theory (de Prado, 2020). The interpretability of models, explainability of results, and use of domain knowledge by the researcher have a central place in the use of machine learning for empirical discovery (Roscher, Bohn, Duarte and Garcke, 2020). Using domain knowledge the researcher has to select the data, specify the appropriate model, and adjust the algorithm to fit certain domain-specific insights and constraints (Roscher, Bohn, Duarte and Garcke, 2020). Furthermore, researchers would still need to apply domain knowledge and build upon existing theory in order to generate an appropriate generalizable representation. A rampant critique against this approach comes in the form that machine learning algorithms are a blackbox and could not be used for scientific analysis. However, as discussed in the previous section, extensive progress has been made on model interpretability and explainability. The interpretability and explainability also heavily depend on the specific algorithms that are used (e.g. intuitive visualisation of tree-based algorithms).

Machine learning can be used for empirical discovery in many ways from clustering to learning efficient data codings via autoencoders. For example, clustering algorithms can be used to classify objects into groups where researchers could further investigate possible commonalities and the drivers of these homogeneities. For example, Huang, Hsu and Chen (2013) examine how decision trees could be used to help characterize and explain clusters. Clustering algorithms can be used for various other tasks in the empirical research process such as discovering population heterogeneity within the sample (Molina and Garip, 2019). As suggested by Israel, Kelly and Moskowitz (2020) machine learning could be used for empirical discovery of financial economic theories by embedding machine learning within a methodological structure according to theoretical constraints. For example Gu, Kelly and Xiu (2019) embed the theoretical considerations of no-arbitrage and the linkage between return and risk in equilibrium in an autoencoder model. Mainly within natural sciences has machine learning been extensively used for generating scientific insights by empirical discovery (Roscher, Bohn, Duarte and Garcke, 2020). Researchers within finance do seem to begin to embrace machine learning for pattern discovery. For example, Leippold, Wang, and Zhou (2020) use machine learning for empirical discovery as they generate systematic signals in the Chinese stock market through the use of a variety of algorithms. They find novel empirical insights which they link with the distinguishing characteristics of the Chinese market. Gu (2020) uses feedforward network models to capture the nonlinear interactions between factors in asset price prediction. Chen (2020) employs an LSTM-GAN architecture to reveal nonlinear interactions between factors and uncover the importance of latent macroeconomic state variables in asset pricing. Bryzgalova, Pelger and Zhu (2019) use pruning of decision trees to select relevant asset pricing portfolios. These studies demonstrate the potential for uncovering nonlinearities outside of the traditional factor representation.

### 4.2 Estimation

Machine learning is particularly useful for processing unstructured data and generating estimates that encapsulate fundamental information contained in this data. The difficulties of handling unstructured (big) data has been one of the main drivers of scientific progress within machine learning. For financial research, one would be mainly interested in processing unstructured data into structured time series variables which could be analysed in statistical models. Unstructured data comes in many different flavors such as language data (text and voice), visual data (pictures and videos), and geo-spatial data. In order to properly process this kind of unstructured data various subfields within machine learning have been developed such as natural language processing (NLP) and computer vision. These subfields generally imply the use of organized methodologies which consist of using machine learning to process unstructured data into something structured that could be analyzed via statistical models or other machine learning algorithms. For example, NLP involves several steps such as cleaning of noise, tokenization, normalization, and vectorization where one converts the data into numeric feature vectors which could be analyzed by machine learning algorithms. Within finance NLP has been extensively used to measure a variety of variables such as annual reports, news articles, central bank policy statements, and corporate earnings calls. Computer vision has been used to analyze things like traffic, ports, and farm land in order to make improved forecasts on a variety of financial instruments and the economy. The processing of geospatial data through machine learning has led to the development of the field of Spatial Finance for which a report of the World Bank (2020) highlighted the ample opportunities.

Machine learning also provides opportunities for improved estimation of structured data because of the more flexible nonlinear patterns that it allows for. In machine learning there are generally less stringent assumptions on the data generating process and relationships relative to conventional statistical methods. As a result, several researchers have used it to quantify new measures or to improve estimation of existing measures. For example, machine learning has been extensively used to generate improved financial risk measures as the task of risk measurement has a strong nonlinear prediction component. As discussed by Gu, Kelly and Xiu (2020) machine learning can be used for improved measurement of risk premia in asset pricing which fundamentally consists of a prediction task. Machine learning could also be used to estimate financial variables that are traditionally based on qualitative human judgment. For example, Ding et al. (2020) show that machine learning methods improve accounting estimates as financial statements sometimes depend on subjective managerial estimates. Moreover, given the limited data availability certain traditional statistical measures could not be estimated in an unbiased way. As previously discussed, machine learning could be used in some cases to provide improved estimates.

### 4.3 Empirical Testing

Machine learning can be used for the empirical testing of theories providing several advantages compared to conventional statistical models. Firstly, machine learning allows for the measurement and testing of more complex non-linear relationships dependencies while providing substantial flexibility in terms of functional forms and models. As argued by Gu, Kelly and Xiu (2019), most leading theoretical asset pricing models predict nonlinear dynamics while linear methods remain the empirical golden standard in the estimation of risk premia. Stevanovic and Surprenant (2019) and Gu, Kelly and Xiu (2020) observe advantages of machine learning methods relative to conventional econometric methods in asset pricing and macroeconomic predictions as a result of these nonlinear interactions. As discussed by de Prado (2019) the closed-form algebraic specifications within conventional econometrics for empirical testing cannot properly handle nonlinearity, discontinuities or topological structures which could be argued to be widely present in financial markets. Secondly, machine learning models generally have less stringent assumptions on the relationships and the distributional properties of the data. Machine learning algorithms allow for the specification that best fits the observations to be flexibly learned in a data-driven fashion rather than relying on the researcher to a priori impose stringent functional specifications. This flexibility decreases the tendency for specification errors in financial research which can lead to false rejection of important variables or relationships (de Prado, 2020). As aforementioned, this comes at the risk of overfit and there are various methodologies developed for this. It should be noted that the imposition of structure based on economic theory does not only aid against overfit, but also enables the testing of hypotheses of financial-economic mechanisms. This is mainly achieved by embedding machine learning within a methodological structure according to theory. The previously discussed studies of Chen, Pelger and Zhu (2020) and Israel, Kelly and Moskowitz (2020) are an excellent example of this. In both studies, the structural formulations and economic constraints not only serve as regularization but also hypothesize economic mechanisms through which the relationship between firm characteristics, macroeconomic variables, and risk premia can be tested via post-hoc feature importance methods. Thirdly, machine learning algorithms are generally better suited in handling high dimensional data which allows for richer models. Especially in asset pricing, machine learning allows researchers to investigate and navigate through the large zoo of factors which have been accumulated during the past decades by analyzing the relationships and evaluating feature importance in a systematic fashion. Machine learning could be used to construct new methodologies for empirical testing which have been traditionally constructed via linear methods presenting elegant algebraic expressions. Bryzgalova, Pelger and Zhu (2019) show that the current empirical methodological approach to construct factors fails to span the stochastic discount factor, suffers from the curse of dimensionality, and contributes to the factor zoo problem. They propose a new machine learning based methodology via decision trees for cross-sectional asset pricing and factor construction that would address the aforementioned problems.

As previously discussed, the process of economically interpretable feature importance analysis lies at the heart of machine learning as an empirical inference tool. The generative formulations in conventional econometric methods lead to a purported ease of testing the statistical significance of underlying hypotheses via p-values. As pointed out in de Prado (2020), p-values rely on strong distributional assumptions which can generate inaccurate results in the presence of collinearity and model misspecification. This is especially problematic in the face of various stylized facts (Conte, 2000) relating to financial returns such as heavy-tailed distribution, gain-loss asymmetry, intermittency, and volatility clustering. OLS regression cannot properly handle nonlinear and high dimensional environments such as we observe in financial markets. As argued by Mullainathan and Spiess (2017), the difficulty of estimating standard errors on the coefficients in machine learning models interferes with how many economists interpret empirical work. Penalized regression methods within machine learning have been used to handle high-dimensional feature sets while keeping the advantages of conventional regression methods (Tibshirani, 2014). Theoretical advancements, such as the properties of estimators, in penalized regression methods make these techniques more direct candidates to be integrated within econometrics (Masini, Medeiros and Mendes, 2020). While Mullainathan and Spiess (2017) are correct in pointing out that an important area of machine learning advancement should be to make econometric sense of the estimated coefficients, developments in interpretability methods have alleviated this. Instead of enforcing intrinsic interpretability by stringent restrictions on the functional form, like is done in conventional econometrics models, many machine learning interpretability methods focus on post-hoc analysis. Post-hoc interpretability aims to allow the researcher to harness the flexible complexity of machine learning algorithms while applying tools to summarize the impact and importance of input features and cross-effects. Deriving statistical significance tests based on asymptotic distributions and simulation studies is a relatively underdeveloped area where econometricians could enrich the empirical testing literature in machine learning. For instance, Horel (2020) approaches the neural network interpretability problem from the perspective of statistical significance testing. They construct a gradient-based test statistic that represents a weighted average of the squared partial derivative of the neural network estimator with respect to a given variable and derive asymptotic distributions. Chen, Pelger and Zhu (2020) apply a similar rationale to assess the relative importance of economic, fundamental, and technical covariates in cross-sectional asset pricing. One can take this a step further and use model-agnostic interpretation methods (such as SHAP) across a suite of models to assess feature importance. Feature importance can be examined across multiple models in order to build consensus and validate the presence of strong common features that drive the empirical process. Moreover it could also be used to uncover nonlinearities and interactions that may have been missed as a result of misspecification of a single model.

Robustness checks of empirical tests are a prevailing practice to examine the validity of the results during different circumstances or under different assumptions. More specifically, it is used to examine the structural validity of the results and robustness of coefficients which are necessary to make valid scientific claims (White and Lu, 2010). Traditional methods for robustness checks in the financial-economic literature include using different control variables, running the analysis on different sub-samples, and using a different model specification. Robustness testing is an implicit procedure in standard machine learning practice to reduce the aforementioned problems of overfit and achieving generalization of the model. Robustness testing in machine learning covers practices such as ensemble, regularization, cross-validation, etc. The machine learning literature provides thus various tools for more extensive robustness testing. One could start with a simpler model and test robustness of the results using machine learning robustness methodologies. As discussed by Mullainathan and Spiess (2017) machine learning could also be used in financial economics to test theories by using machine-learning-based benchmarks to test how well theories are empirically performing. Researchers could thus compare the results obtained by a simpler model to the dynamics created by a more complex data-driven model. In this way, one could get a better sense of the merits of the conclusions of the theory in question. Finally, and as previously discussed, the in-sample assessment methodology within conventional econometrics encourages p-hacking through multiple testing. Applying machine learning methodologies and the philosophy of out-of-sample generalizability could help combat p-hacking in financial research.

### 4.4 Causal Inference

There is an inherent difficulty of causality analysis in financial-economic problems because controlled experiments are in most cases not possible and the researcher is just an observing party. Moreover, as discussed by Xu (2018), causal inference in finance poses an epistemological problem as it is rather unclear how to properly define numerically causality in financial time series. The inference of causality tends to be an arduous task with traditional econometric methods as it requires, for example, instrumental variables or an appropriate independent treatment. Conventional econometric methods all tend to use observational data, and as a result, we argue that machine learning methods for causal inference could provide researchers with an expanded toolbox to examine causality. Machine learning could be used for causal inference in two ways, namely, improve traditional methods in causal inference and introduce new methodologies for causality analysis.

Varian (2016) discusses how machine learning can be used to improve traditional methodologies of causal inference in economics as he argues that these fundamentally depend on a prediction model. Tiffin (2019) also elaborates on how machine learning could be introduced to solve causal inference problems as these can be broken into prediction problems. Machine learning algorithms, with their strong predictive ability, could be used to estimate the counterfactual in natural experimental designs. Firstly, machine learning could be used to improve techniques like regression discontinuity, instrumental variables and difference-in-differences as these methods would benefit from an improved and more flexible (non-linear) prediction. In these cases, a prediction problem is hidden within methodology to solve and machine learning models could be used to boost the predictive performance. For instance, for the instrumental variables problem, Belloni, Chen, Chernozhukov, and Hansena (2012) have used LASSO regression to determine the optimal set of instruments in the first stage. Varian (2016) explains how regression discontinuity can be treated as two separate predictions via a regression below and above a threshold.

Secondly, machine learning offers new ways of addressing causal inference, especially in the context of time series. Tiffin (2019) proposes to leverage the properties of random forests to estimate the so-called feature importances, in order to isolate the causal factors. For each tree of the forest, half of the data is used for average effect estimation and half for variable selection. This enables to efficiently determine the causal factors and quantify the varying effects on the population while controlling for the variance of the model. In a similar fashion, Athey and Imbens (2016) developed a causal tree to account for heterogeneous effects on individuals. The causal tree partitions the space of individuals and estimates a different effect for these various groups. Pearl (2000) formalized a framework of causal inference graphs where causal factors are defined as variables that could impact the output even when conditioned on any other variable. The approach gets rid of spurious correlations as variables that have numerical correlations with the variable of interest are discarded if the numerical correlation disappears when conditioned on another variable. In a similar manner, Bontempi and Flauder (2015) have used the asymmetry between conditional dependencies to extract causal information. In their approach, machine learning is used to estimate a measure of the degree of asymmetry as a function of causal factors. Another method which has been used in social science is structural equation modeling where a priori assumptions about the causal relationships are made. Researchers define a structural model to link latent variables via a system of simultaneous equations and a measurement model to define these latent variables in terms of factors (Kaplan, 2008). Machine learning has been used to augment structural equation models through Bayesian Causal Networks which combine structural equation modeling with graph models which represent the dependency structure. This methodology allows for counterfactual analysis and the dependency graphs allow researchers to apply domain expertise (QuantumBlack, 2019). Researchers can update the beliefs based on domain knowledge and retrain the Bayesian network in a feedback loop fashion. Another important advancement is Bayesian Time Series Models (BTSM) which can be used for causal inference in time series (Brodersen, Gallusser, Koehler, Remy, and Scott, 2015). BTSM essentially is a machine learning based state-space model for time series and allows for feature selection in Bayesian sense (Jammalamadaka, Qiu and Ning, 2018). It processes multiple time series at the same time, and the correlation between the time series imparts robustness to the model. The underlying idea is to leverage the signal in the "control" time series that should not be impacted by the cause whose effects are to be inferred. Causal inference is then performed taking into account the correlation of the control time series with the "response" time series that are impacted. One gets rid of spurious effects which are changes in the underlying process independent of the considered cause.

### 4.5 Prediction

Prediction holds a central place within empirical financial research as capital allocation decisions require judgments on uncertain future states, rewards and risks. Asset pricing essentially encompasses the estimation of risk premia, which represent expected returns of risky assets, and thus involves a substantial prediction component (Gu, Kelly and Xiu, 2019). However, as discussed by de Prado (2019) and Mullainathan and Spies (2017), the rationale behind the traditional econometric approach is rather incompatible with out-of-sample prediction. While a large part of the machine learning literature focuses on predictive learning and there is a general emphasis on out-of-sample predictive performance. This emphasis on out-of-sample prediction performance lies in the view that if a model describes the data well it should generalize equally well and little assumptions on the data generating process are needed (Rudin, 2015). Machine learning mainly provides opportunities for improved prediction in the case of nonlinearities, higher-order interactions, and environments with large amounts of correlated features (Athey and Imbens, 2019). Nonlinear tendencies and higher-order interactions in financial relationships should be rather unsurprising given some of the stylized facts of returns such as heavy-tailed distribution, gain-loss asymmetry, and volatility clustering (Conte, 2001). These could be theoretically justified via market inefficiencies and imperfections. Especially the ability to take into account large (changing) sets of predictors has been found helpful within financial markets (Chinco, Clark-Joseph and Ye, 2017). For these reasons machine learning tends to outperform in prediction tasks compared to conventional econometric methods. For example, penalized regression models have been used as high-dimensional alternatives to OLS for financial predictions (Athey and Imbens, 2019). Nonlinear models, such as decision trees and neural networks, have generated improved prediction results because of nonlinear interactions due to market imperfections (Masini, Medeiros and Mendes, 2020). This outperformance of machine learning in forecasting has been observed by among others Gu, Kelly and Xiu (2020) for asset pricing and Stevanovic and Surprenant (2019) for macroeconomics. This tendency for outperformance can be especially helpful for tasks, such as asset pricing, where the quality of the predictions are an important determinant to understand the financial mechanisms and test theories. Furthermore, an increasingly popular application of predictive models in economics and finance is nowcasting. Nowcasting has been traditionally done through regression or factor models (Higgings, 2014). Machine learning could be useful for nowcasting as it requires a large set of dynamic contemporaneous predictors and new unstructured data sources provide promising opportunities to generate nowcasts. For example, Varian (2014) argues that Bayesian Time Series Models are useful for nowcasting since they allow for direct updating according to contemporaneous features and allow for time series forecasting with a large amount of features (Scott and Varian, 2014).

Appropriate model selection, feature selection, managing overfit, model interpretability, ensuring robustness are all necessary steps in the prediction exercise. As discussed in previous sections, these elements are especially challenging in finance. This illustrates that the various tasks within the empirical process are interconnected and one needs an unified financial machine learning framework. Recently there has been progress in automated machine learning (AutoML) which attempts to automatize the aforementioned process of building a prediction model. For example, Google (2020) recently launched a AutoML time series forecasting model which automates the feature selection, model selection, and (hyper)parameter tuning process. However, researchers should be careful when using these AutoML models for financial research considerations as these are usually not fitted for the specific data environment in finance. Related to this, is the previously mentioned argument that predictions can be improved by embedding financial theory into machine learning models. Domain knowledge by researchers and a priori theoretical considerations are key in building appropriate financial machine learning models for prediction. This, again, shows the need for a branch of financial machine learning to develop and gain econometric validation.

## 5. Conclusion

Machine learning algorithms have been mainly developed for particular data environments, and there are some key differences with financial markets in this regard. Not only do difficulties arise due to some of the peculiar features of financial markets, there is a fundamental tension between the underlying paradigm of machine learning and the research philosophy in financial economics. Despite some challenges, machine learning can be unified with financial research when appropriate methodological adjustments are made. Financial machine learning provides a plethora of opportunities for financial research to supplement or complement some conventional econometric methodologies. More specifically, it can be used for various parts of the research process such as data pre-processing, estimation, empirical discovery, testing, causal inference and prediction. However, econometrics has not caught up with the progress in statistical learning and the empirical researcher's toolbox has not changed much in the past decade. As a result there are substantial advances to be made in the cross-section of econometrics and machine learning, and how to unify these into a single empirical framework. Applying machine learning in a scientifically sound manner requires fundamental theoretical reasoning to construct the appropriate empirical setting and examine the soundness of the results. To account for the idiosyncrasies of financial markets and the specific needs of financial researchers, it is important to develop the branch of financial machine learning. Econometrics has been a major contribution in the development of time series analysis in statistics, and our hope is that it can have a similar impact in some areas within machine learning as well.

# 6. References

Anandakrishnan, A., Kumar, S., Statnikov, Faruquie, A. T., and D. Xu. 2018. Anomaly detection in finance: editors' introduction. *KDD 2017 Workshop on Anomaly Detection in Finance*: 1-7.

Arnott, R., Harvey, C. R., and H. Markowitz. 2019. A backtesting protocol in the era of machine learning. *The Journal of Financial Data Science* 1: 64-74.

Assefa, S., Dervovic, D., Mahfouz, M., Balch, T., P. Reddy and M. Veloso. 2019. Generating synthetic data in finance: opportunities, challenges and pitfalls. *NeurIPS Workshop on Robust AI in Financial Services.*

Athey S. and Imbens G.W. 2016. Recursive partitioning for heterogeneous causal effects. *Proceedings of the National Academy of Science* 113 :7353-7360.

Athey, S., and Imbens, G.W. 2017. The state of applied econometrics: Causality and policy evaluation. *Journal of Economic Perspectives,* 31(2), pp.3-32.

Athey, S., and Imbens, G. W. 2019. Machine learning methods economists should know about. *Annual Review of Economics* 11: 685-725.

Bailey, D.H., Borwein, J., Lopez de Prado, M. and Zhu, Q.J., 2016. The probability of backtest overfitting. *Journal of Computational Finance*.

Bailey, D.H. and De Prado, M.L., 2014. The deflated Sharpe ratio: correcting for selection bias, backtest overfitting, and non-normality. *The Journal of Portfolio Management*, 40(5), pp.94-107.

Balcerak M. and T. Schmelwer. 2020. "Constructing trading strategy ensembles by classifying market states." Available at arXiv preprint arXiv:2012.03078.

Bauwens, L., Preminger, A. and J. Rombouts. 2006. "Regime Switching GARCH Models" Available at SSRN: https://ssrn.com/abstract=914144.

Belloni A., Chen D., Chernozhukov V., and C. Hansen. 2012. Sparse Models and Methods for Optimal Instruments With an Application to Eminent Domain. *Econometrica, Journal of the Econometric Society* 80: 2369-2429.

Biddle, J. 2017. Statistical Inference in Economics, 1920–1965: Changes in Meaning and Practice. *Journal of the History of Economic Thought* 39: 149-173.

Bildirici, M. and Ersin, Ö., 2016. Markov Switching Artificial Neural Networks for Modelling and Forecasting Volatility: An Application to Gold Market. *Procedia economics and finance*, *38*, pp.106-121.

Black, F. 1986. Noise. *The Journal of Finance* 41: 528-543.

Boiko Ferreira, L. E., Barddal J. P., Gomes H. M and F. Enembreck. 2018. An Experimental Perspective on Sampling Methods for Imbalanced Learning From Financial Databases. *International Joint Conference on Neural Networks (IJCNN 2018)*

Bontempi, G, Ben Taieb, S and Y.-A. Le Borgne. 2013. Machine learning strategies for time series forecasting. *Second European Business Intelligence Summer School.*

Bontempi, G. and Flauder, M., 2015. From dependency to causality: a machine learning approach. *The Journal of Machine Learning Research* 16: 2437-2457.

Box, G.E., 1979. Robustness in the strategy of scientific model building. *Robustness in statistics*: 201-236.

Breiman, L. 2001. Statistical modeling: The two cultures (with comments and a rejoinder by the author). *Statistical science* 16: 199-231.

Brodersen, K. H., Gallusser, F., Koehler, J., Remy, N., and L. S. Scott. 2015. Inferring causal impact using Bayesian structural time-series models. *The Annals of Applied Statistics* 9: 247-274.

Bryzgalova, S., Pelger, M., and J. Zhu. 2019. "Forest through the trees: Building cross-sections of stock returns." Available at SSRN 3493458.

Burkov, A. 2019. *The Hundred-Page Machine Learning Book*. ISBN-13: 978-1999579500.

Bzdok, D., Altman, N. and M. Krzywinski. 2018. Statistics versus machine learning. *Nature Methods* 15: 233–234.

Cai, J., Luo, J., Wang, S. and S. Yang. 2018. Feature selection in machine learning: A new perspective. *Neurocomputing* 300: 70-79.

Cao L. and Q. Gu. 2002. Dynamic support vector machines for non-stationary time series forecasting. *Intelligent Data Analysis* 6:67-83

Cerqueira, V., Torgo, L. and C. Soares. 2019. "Machine learning vs statistical methods for time series forecasting: Size matters." Available at arXiv preprint arXiv:1909.13316.

Cerqueira, V., Torgo, L. and I. Mozetič. 2020. Evaluating time series forecasting models: An empirical study on performance estimation methods. *Machine Learning* 109: 1997-2028.

Chandar, S.K., Sumathi, M. and Sivanandam, S.N., 2016. Prediction of stock market price using hybrid of wavelet transform and artificial neural network. *Indian journal of Science and Technology*, 9(8), pp.1-5.

Charpentier, A., Elie, R., and C. Remlinger. 2020. "Reinforcement Learning in Economics and Finance." Available at arXiv preprint arXiv:2003.10014.

Chen, L., Pelger, M. and J. Zhu. 2020. "Deep Learning Asset Pricing" Available at SSRN: https://ssrn.com/abstract=3350138.

Cheng C., Sa-Ngasoongsong A., Beyca O., Le T., Yang H., Kong Z. and S. T. S. Bukkapatnam. 2015. Time series forecasting for nonlinear and non-stationary processes: a review and comparative study. *IIE Transactions* 47: 1053-1071.

Cont, R. 2001. Empirical properties of asset returns: stylized facts and statistical issues. *Quantitative Finance* 1: 223-236.

Crone, S.F., Hibon, M. and Nikolopoulos, K., 2011. Advances in forecasting with neural networks? Empirical evidence from the NN3 competition on time series prediction. *International Journal of forecasting*, 27(3), pp.635-660.

Dernoncourt, D., Hanczar, B. and J.D. Zucker. 2014. Analysis of feature selection stability on high dimension and small sample data. *Computational statistics & data analysis* 71: 681-693.

de Prado, M. L. 2018. *Advances in financial machine learning.* Hoboken, NJ: John Wiley & Sons.

de Prado, M. L. 2019. "Beyond Econometrics: A Roadmap Towards Financial Machine Learning." Available at SSRN 3365282.

de Prado, M. L. 2016. Building diversified portfolios that outperform out of sample. *The Journal of Portfolio Management* 42: 59-69.

de Prado, M. L. 2020. *Machine Learning for Asset Managers.* Cambridge, UK: Cambridge University Press.

Dundar, M., Krishnapuram, B., Bi, J., and R. B. Rao. 2007. Learning Classifiers When the Training Data Is Not IID. *Proceedings of the 20th International Joint Conference on Artificial Intelligence*: 756-761.

Fan, J., Lv, J. and L. Qi. 2011. Sparse high-dimensional models in economics. *Annual Review of Economics* 3: 291-317.

Fan, J. and J. Lv. 2010. A selective overview of variable selection in high dimensional feature space. *Statistica Sinica* 20: 101.
Feng, J. and Simon, N., 2018. Gradient-based regularization parameter selection for problems with nonsmooth penalty functions. *Journal of Computational and Graphical Statistics*, 27(2), pp.426-435.

Fernandez C., and M. Steel. 1998. On Bayesian Modeling of Fat Tails and Skewness. *Journal of the American Statistical Association* 93: 359-371.

Giles C. L., Lawrence S. and A. C. Tsoi. 2001. Noisy Time Series Prediction using Recurrent Neural Networks and Grammatical Inference. *Machine Learning* 44: 161–183.

Gu, S., Kelly, B., and D. Xiu. 2019. Autoencoder Asset Pricing Models. *Yale ICF Working Paper No*. 2019-04, *Chicago Booth Research Paper* No. 19-24

Gu, S., Kelly, B., and D. Xiu. 2020. Empirical asset pricing via machine learning. *The Review of Financial Studies* 33: 2223-2273.

Guidolin, M. 2011. Markov Switching Models in Empirical Finance. *Advances in Econometrics* 27.

Gupta, S., and A. Gupta. 2019. Dealing with Noise Problem in Machine Learning Data-sets: A Systematic Review. *Procedia Computer Science* 161: 466-474.

Hanneke, S. 2016. The optimal sample complexity of PAC learning. *The Journal of Machine Learning Research* 17: 1319-1333.

Harvey, C.R. and Y. Liu. 2019. "A census of the factor zoo." Available at SSRN 3341728.

Harvey, C.R., 2017. Presidential address: The scientific outlook in financial economics. *The Journal of Finance*, 72(4), pp.1399-1440.
28

Hastie, T., Tibshirani, R. and M. Wainwright. 2015. *Statistical learning with sparsity: the lasso and generalizations:* CRC press.

He Q. Q., Pang P.C. I. and Y. W. Si. 2019. Transfer Learning for Financial Time Series Forecasting. *Pacific Rim International Conference on Artificial Intelligence 2019.*

Hey, T. 2009. *The fourth paradigm: data-intensive scientific discovery*, vol. 1: Microsoft research.

Higgins, Patrick C. 2014. "GDPNow: A Model for GDP 'Nowcasting'" FRB Atlanta Working Paper. Available at SSRN: https://ssrn.com/abstract=2580350.

Holmes M. J. and B. Silverstone. 2007. "Business Confidence and Cyclical Turning Points: A Markov-Switching Approach," Working Papers in Economics, University of Waikato.

Horel, E., and K. Giesecke. 2020. Significance Tests for Neural Networks. *Journal of Machine Learning Research* 21: 1-29.

Huang, T.C.K., Hsu, W.H. and Y.L. Chen. 2013. Conjecturable knowledge discovery: A fuzzy clustering approach. *Fuzzy Sets and Systems* 221: 1-23.

Huang, S.C. and Wu, T.K., 2010. Integrating recurrent SOM with wavelet-based kernel partial least square regressions for financial forecasting. *Expert Systems with Applications*, 37(8), pp.5698-5705.

Ilhan, F., Karaahmetoglu, O., Balaban, I. and Kozat, S.S., 2020. Markovian RNN: An Adaptive Time Series Prediction Network with HMM-based Switching for Nonstationary Environments. *arXiv preprint arXiv:2006.10119*.

Islam, S. R., Ghafoor, S. K., and W. Eberle. 2018. Mining illegal insider trading of stocks: A proactive approach. *2018 IEEE International Conference on Big Data*: 1397-1406

Israel, R., Kelly, B. T., and T. J. Moskowitz. 2020. "Can Machines' Learn Finance?." Available at SSRN 3624052.

Jain, A. and D. Zongker. 1997. Feature selection: Evaluation, application, and small sample performance. *IEEE transactions on pattern analysis and machine intelligence* 19: 153-158.

Jeong, G., and H. Y. Kim. 2019. Improving financial trading decisions using deep Q-learning: Predicting the number of shares, action strategies, and transfer learning. *Expert Systems with Applications* 117: 125-138,

Jordan, M. I., and T. M. Mitchell. 2015. Machine learning: Trends, perspectives, and prospects. *Science* 349: 255-260.

Kaplan, D. 2008. *Structural equation modeling: Foundations and extensions*, vol. 10: Sage Publications.

König, G., Molnar, C., Bischl, B. and Grosse-Wentrup, M., 2020. "Relative Feature Importance." Available at arXiv preprint arXiv:2007.08283.

Koshiyama, A. and Firoozye, N., 2019. Avoiding Backtesting Overfitting by Covariance-Penalties: an empirical investigation of the ordinary and total least squares cases. *The Journal of Financial Data Science*, 1(4), pp.63-83.

Kumar, I.E., Venkatasubramanian, S., Scheidegger, C. and Friedler, S., 2020. Problems with Shapley-value-based explanations as feature importance measures. *International Conference on Machine Learning*: 5491-5500.

Lahmiri, S., 2014. Comparative study of ECG signal denoising by wavelet thresholding in empirical and variational mode decomposition domains. *Healthcare technology letters*, 1(3), pp.104-109.

Langley, P. and Zytkow, J.M., 1989. Data-driven approaches to empirical discovery. *Artificial Intelligence* 40:.283-312.

Leippold, M.,Wang, Q., and Zhou, W. 2020. "Machine-Learning in the Chinese Factor Zoo." Available at SSRN: https://papers.ssrn.com/sol3/papers.cfm?abstract_id=3754339

Leo, M, Sharma, S., and K. Maddulety. 2019. Machine learning in banking risk management: A literature review. *Risks* 7: 29.

Li, A., Wu, J., and Z. Liu. 2017. Market manipulation detection based on classification methods. *Procedia Computer Science* 122: 788-795.

Liao, J. J., Shih, C. H., Chen, T. F. and M. F. Hsu. 2014. An ensemble-based model for two-class imbalanced financial problem. *Economic Modelling* 37: 175-183.

Lin, W., Hu Y. and C. Tsai. 2011. Machine Learning in Financial Crisis Prediction: A Survey. *IEEE Transactions on Systems, Man, and Cybernetics* 42 : 421-436.

Liu, B., Wei, Y., Zhang, Y. and Q. Yang. 2017. Deep Neural Networks for High Dimension, Low Sample Size Data. *IJCAI* : 2287-2295.

Lu, X., and H. White. 2014. Robustness checks and robustness tests in applied economics. *Journal of Econometrics* 178: 194-206.

Lundberg, S., and S. Lee. 2017. A Unified Approach to Interpreting Model Predictions. *31$^{st}$ Conference on Neural Information Processing Systems*.

Ma, S. and Tourani, R., 2020. Predictive and causal implications of using shapley value for model interpretation. *Proceedings of the 2020 KDD Workshop on Causal Discovery*: 23-38.

Makridakis, S., Spiliotis, E. and Assimakopoulos, V., 2018. Statistical and Machine Learning forecasting methods: Concerns and ways forward. *PloS ONE* 13.

McLean, R. D., and J. Pontiff. 2016. Does academic research destroy stock return predictability? *The Journal of Finance* 71: 5-32.

Melkumova, L.E. and Shatskikh, S.Y., 2017. Comparing Ridge and LASSO estimators for data analysis. *Procedia engineering*, 201, pp.746-755.

Merton, R. C. 1980. "On estimating the expected return on the market: An exploratory investigation." Working paper No. w0444, National Bureau of Economic Research.

Molina, M., and F. Garip. 2019. Machine learning for sociology. *Annual Review of Sociology* 45: 27-45.

Molnar, C. 2019. "Interpretable Machine Learning: A Guide for Making Black Box Models Explainable." Available at https://christophm.github.io/interpretable-ml-book/.

Molnar, C., König, G., Herbinger, J., Freiesleben, T., Dandl, S., Scholbeck, C.A., Casalicchio, G., Grosse-Wentrup, M. and Bischl, B. 2020. "Pitfalls to avoid when interpreting machine learning models." Available at arXiv preprint arXiv:2007.04131.

Mullainathan, S., and J. Spiess. 2017. Machine learning: an applied econometric approach. *Journal of Economic Perspectives* 3: 87-106.

Muthukrishnan, R. and Rohini, R., 2016, October. LASSO: a feature selection technique in predictive modeling for machine learning. *IEEE international conference on advances in computer applications* (ICACA) (pp. 18-20). IEEE.

Nori, H., Jenkins, S., Koch, P., and R. Caruana. 2019 "InterpretML: A Unified Framework for Machine Learning Interpretability" Microsoft Corporation.

Patterson, D.J., Ariel, Y., Burks, B., Gratcheva, E.M., Hosking, J.S., Klein, N., Protopopova, L., Ruiz, E.E.T., Schmitt, S., Shackelton, B. and D. Wang. 2020. "Spatial finance: Challenges and opportunities in a changing world." World Bank. Available at http://hdl.handle.net/10986/34894.

Pearl, J. 2000. *Causality: Models, Reasoning, and Inference.* Cambridge, UK: Cambridge University Press.

Peters, J., Janzing, D. and Schölkopf, B., 2017. *Elements of causal inference*, p. 288, Cambridge, MA: MIT Press.

Philips, P. C. B. 1995. Nonstationary time series and cointegration. *Journal of Applied Econometrics* 10: 87-94.

Piger J. 2009. Econometrics: Models of Regime Changes. *Complex Systems in Finance and Econometrics*, New York, NY: Springer.

Pohl, W., Schmedders, K., and O. Wilms. 2018. Higher order effects in asset pricing models with long-run risks. *The Journal of Finance* 73: 1061-1111.

Primiceri, G.E., Lenza, M. and D. Giannone. 2018. "Economic Predictions with Big Data: The Illusion of Sparsity." Federal Reserve Bank of New York. Available at SSRN: https://ssrn.com/abstract=3166281.

Qiu, J., Jammalamadaka, S. R. and N. Ning. 2018. Multivariate Bayesian Structural Time Series Model. *Journal of Machine Learning Research* 19: 1-33.

Rudin, C. 2015. "Can Machine Learning Be Useful for Social Science." *The Cities: An essay collection from the Decent City initiative*, vol. 9: 86-90.

Shalizi C.R., Jacobs A. Z., Klinkner K. L., and A. Clauset. 2011."Adapting to Non-stationarity with Growing Expert Ensembles." Available at arXiv preprint arXiv:1103.0949.

Shrikumar, A., Greenside, P., and A. Kundaje. 2019. Learning Important Features Through Propagating Activation Differences. *Proceedings of the 34th International Conference on Machine Learning*: 3145-3153.

Sirignano, J.A., 2019. Deep learning for limit order books. *Quantitative Finance*, 19(4), pp.549-570.

Steinbach, M., Ertöz, L. and V. Kumar. 2004. The challenges of clustering high dimensional data. *New directions in statistical physics*, p. 273-309, Berlin, Heidelberg: Springer.

Sundararajan, M. and Najmi, A., 2020. The many Shapley values for model explanation. *International Conference on Machine Learning*: 9269-9278.

Sugiyama, M. and M. Kawanabe. 2012. *Machine Learning in Non-Stationary Environments: Introduction to Covariate Shift Adaptation,* Cambridge, MA: MIT Press.

Taleb, N. 2020. *Statistical Consequences of Fat Tails: Real world preasymptotics, epistemology and applications*. STEM Academic Press.

Taylor S.J. and B. Letham. 2018. Forecasting at scale. *The American Statistician* 72: 37-45.

Tibshirani, R. 2014. *High-dimensional regression: Lecture notes Advanced Methods for Data Analysis*. Carnegie Mellon University.

Tiffin, A. 2019. "Machine Learning and Causality: The Impact of Financial Crises on Growth" IMF Working Paper No. 19/228.

Vabalas, A., Gowen, E., Poliakoff, E. and A. J. Casson. 2019. Machine learning algorithm validation with a limited sample size. *PloS ONE* 14.

van der Ploeg, T., Austin, P. C. and E. W. Steyerberg. 2014. Modern modelling techniques are data hungry: a simulation study for predicting dichotomous endpoints. *BMC medical research methodology* 14: 137.

Vapnik, V. 2013. *The Nature of Statistical Learning Theory*. Springer Science & Business Media.

Varian, H. R. 2014. Big data: New tricks for econometrics. *Journal of Economic Perspectives* 28: 3-28.

Varian, H. R. 2016. Causal inference in economics and marketing. *Proceedings of the National Academy of Sciences* 113: 7310-7315.

Varma, S. and Simon, R., 2006. Bias in error estimation when using cross-validation for model selection. *BMC bioinformatics*, *7*(1), pp.1-8.

Vaughan, J., Sudjianto, A., Brahimi, E., Chen, J. and V. N. Nair. 2018. "Explainable Neural Networks based on Additive Index Models", Corporate Model Risk, Wells Fargo.

Verleysen, M. and D. François. 2005. The curse of dimensionality in data mining and time series prediction. *International work-conference on artificial neural networks,* p. 758-770, Berlin, Heidelberg: Springer.

Wong, J.C., 2020. "Computational causal inference." arXiv preprint arXiv:2007.10979.

Xu, L., 2018. Machine learning and causal analyses for modeling financial and economic data. *Applied Informatics* 5: 1-42.

Yazdani, A. 2020. Machine Learning Prediction of Recessions: An Imbalanced Classification Approach. *The Journal of Financial Data Science* 2: 21-32;

Yoon, J., Jarrett, D. and M. van der Schaar. 2019. Time-series Generative Adversarial Networks. *Advances in Neural Information Processing Systems* 32.

Zhao, Z., Rao, R., Tu, S. and Shi, J., 2017, November. Time-weighted LSTM model with redefined labeling for stock trend prediction. In 2017 *IEEE 29th international conference on tools with artificial intelligence (ICTAI)* (pp. 1210-1217). *IEEE*.